\documentclass[twocolumn,superscriptaddress,pra,letterpaper,showpacs]{revtex4}

\usepackage{hyperref}
\usepackage{graphicx}
\usepackage{rotating}
\usepackage{amsmath}
\usepackage{amsfonts}
\usepackage{amssymb}
\usepackage{enumerate}
\usepackage{longtable}
\usepackage{dcolumn}
\usepackage{bbm}

\begin{document}

\title{Rigorous Bounds for Optimal Dynamical Decoupling}

\author{G\"otz S. Uhrig}
\email{goetz.uhrig@tu-dortmund.de}
\affiliation{Lehrstuhl f\"{u}r Theoretische Physik I, 
Technische Universit\"{a}t
Dortmund, Otto-Hahn Stra\ss {}e 4, 44221 Dortmund, Germany}

\author{Daniel A. Lidar}
\email{lidar@usc.edu}
\affiliation{Departments of Chemistry, Electrical Engineering, and Physics, 
Center for
Quantum Information \& Technology, University of Southern California, Los
Angeles, California 90089, USA}

\date{\textrm{\today}}

\begin{abstract}
We present rigorous performance bounds for the optimal dynamical
decoupling pulse sequence protecting a quantum bit (qubit) 
against pure dephasing. Our
bounds apply under the assumption of instantaneous pulses and of bounded
perturbing environment and qubit-environment Hamiltonians. We show
that if the total sequence time is fixed 
the optimal sequence can be used to make the
distance between the protected and unperturbed qubit
states arbitrarily small in the number of applied pulses.
If, on the other hand, the minimum pulse interval is
fixed and the total sequence time is allowed to scale with the number of 
pulses, then longer sequences need not always be advantageous.
The rigorous bound may serve as testbed for
approximate treatments of optimal decoupling in
bounded models of decoherence.
\end{abstract}

\pacs{03.67.Pp, 82.56.Jn, 76.60.Lz, 03.65.Yz}
\maketitle

\section{Introduction}

Quantum systems tend to rapidly decohere due to the coupling to their
environments, a process which is especially detrimental to quantum
information processing and high resolution spectroscopy \cite{Breuer:book}. 
Of the many methods which have been proposed in
recent years to overcome the damage caused by decoherence, we focus here on
dynamical decoupling (DD), a method for suppressing decoherence whose
origins can be traced to the Hahn spin echo \cite{Hahn:50}. In DD, one
applies a series of strong and frequent pulses to a system, designed to
decouple it from its environment
\cite{Viola:98,Ban:98,Zanardi:98b,Viola:99}. Recently, it was
discovered how to optimally suppress decoherence of 
a single qubit using DD, under the idealization of instantaneous
pulses \cite{Uhrig:07,WFL:09,Pasini:09}. One of us found an optimal pulse
sequence (later dubbed 
Uhrig DD, or UDD) for suppressing pure dephasing (single-axis decoherence)
of a qubit coupled to a boson bath with a hard frequency cut-off
\cite{Uhrig:07}. In UDD one applies a series of $N$ instantaneous
$\pi $ pulses at instants $t_{j}$ ($j\in \{1,2\ldots N\}$), with the instants 
given by $t_{j}=T\delta _{j}$ where $T$ is the total time of the sequence and 
\begin{equation}
\delta _{j}=\sin ^{2}(j\pi /(2N+2)).  
\label{udd}
\end{equation}
By optimal it is meant that with each additional pulse the sequence
suppresses dephasing in one additional order in an expansion in $T$, i.e., 
$N$ pulses reduce dephasing to $\mathcal{O}(T^{N+1})$. The existence and
convergence of an expansion in powers of $T$, at least as an asymptotic
expansion, is a necessary assumption \cite{Yang:08,Pasini:09}.

The UDD sequence was first conjectured \cite{lee:160505,Uhrig:08} and then
proven \cite{Yang:08} to be universal, in the sense that it applies to any
bath causing pure dephasing of a qubit, not just bosonic baths. The
performance of the UDD sequence was tested, and its advantages over other
pulse sequences confirmed under appropriate circumstances, in a series of
recent experiments \cite{Biercuk:09,biercuk:062324,Du:09}. Its limitations
as a function of sharpness of the bath spectral density high frequency
cut-off \cite{Pasini:09a} and as a function of timing constraints
\cite{Hodgson:09} have also been discussed. 

In order to suppress general (three-axis) decoherence on a qubit
to all orders
concatenated sequences are needed \cite{Khodjasteh:05,KhodjastehLidar:07},
whose efficiency can be
improved by using UDD building blocks \cite{Uhrig:09b}.
A near optimum suppression  is achieved by quadratic dynamic
decoupling (dubbed QDD). This scheme was proposed
and numerically tested in Ref.\ \cite{WFL:09} and analytically
corroborated in Ref.\ \cite{Pasini:09}. 
In QDD, a sequence of $(N+1)^{2}$
pulses comprising two nested UDD sequences suppresses general qubit
decoherence to $\mathcal{O}(T^{N+1})$, which is known from brute-force
symbolic algebra solutions for small $N$ to be near-optimal \cite{WFL:09}.

While rigorous performance bounds have been derived previously for 
periodic and concatenated DD pulse
sequences \cite{KhodjastehLidar:08,LZK:08,NLP:09}, no such
performance bounds have yet been derived for optimal decoupling pulse
sequences, in particular UDD and QDD. In this
work we focus on UDD and obtain rigorous performance bounds.
We postpone the problem of finding rigorous QDD performance bounds to a
future publication. Our main result here is an analytical upper bound for
the distance between UDD-protected states subject to pure
dephasing and unperturbed states, as a function of the natural dimensionless 
parameters of the
problem, namely the total evolution time $T$ measured in units of the
maximal intra-bath energy $J_0$,
and in units of the system-bath coupling strength $J_z$.
The bound shows that this distance
(technically, the trace-norm distance), can be made arbitrarily small as a
function of the number of pulses $N$, as 
$(1/N!)(J_0 T+ J_z T)^{N}$.
This presumes that the bounds $J_\alpha$ ($\alpha\in\{0,z\}$) are finite, an 
assumption which will fail for unbounded
baths, such as oscillator baths. In such cases, which includes the
ubiquituous spin-boson model, our analysis does not apply. Alternative
approaches, such as those based on correlation function bounds 
\cite{NLP:09}, are then required.

We begin by introducing the model of pure dephasing in the presence of
instantaneous DD pulses in Section \ref{sec:model}. We then derive a
general time-evolution bound in Section \ref{sec:t-bound}, without reference
to any particular pulse sequence. In Section \ref{sec:dep-bound} we
specialize this bound to the UDD sequence. Then, in Section
\ref{sec:D-bound}, we obtain our main result:\ an upper bound on the
trace-norm distance between the UDD-protected and unperturbed
states. In Section \ref{analysis} we analyze the implications of this
bound in the more realistic setting when only a certain minimal interval
between consecutive pulses can be attained.  Certain technical
details are presented in the Appendix, including the first complete
universality proof of the UDD sequence, which does not rely on the
interaction picture.

\section{Model}
\label{sec:model}

We start from the general, uncontrolled, time-independent system-bath
Hamiltonian for pure dephasing 
\begin{equation}
H_{\mathrm{unc}}=I_{\mathrm{S}}\otimes B_0
+\sigma _{z}\otimes B_z,
\label{hamil1}
\end{equation}
where $B_0$ and 
$B_z$
are bounded but otherwise arbitrary operators
acting on the bath Hilbert space $\mathcal{H}_{\mathrm{B}}$, $I_{\mathrm{S}}$
is the identity operator on the system Hilbert space 
$\mathcal{H}_{\mathrm{S}}$, and $\sigma _{z}$ is the diagonal Pauli matrix. 
The bath operator  $B_z$ need not be traceless, i.e., 
we allow for the possibility of a pure-system term 
$\sigma _{z}\otimes I_{\mathrm{B}}$ in the system-bath interaction term 
$H_{\mathrm{SB}}:= \sigma _{z}\otimes B_z$. 
Such internal system dynamics will be
removed by the DD pulse sequence we shall add next, along with the coupling
to the bath. However, the assumption of pure dephasing means that we assume
that the level splitting of the system, i.e., any term proportional to 
$\sigma _{\bot }$ (with $\sigma _{\bot }$ being $\cos(\varphi)
\sigma_x+\sin(\varphi)\sigma_y$ for arbitrary $\varphi$)
acting on the system, is fully controllable. Otherwise the
model is one of general decoherence, and our methods require a modification
along the lines of Refs.~\cite{WFL:09} and \cite{Pasini:09}. 
If the system described by Eq.~\eqref{hamil1} is subject to $N$ 
instantaneous $\pi $ pulses at the instants 
$\{t_{j}:=T\delta _{j}\}_{j=1}^{N}$ about a
spin axis perpendicular to the $z$-axis, i.e., if the Hamiltonian 
$H_{\mathrm{DD}}(t)=\frac{\pi }{2}\sum_{j=1}^{N}\delta (t-t_{j})\sigma _{\bot
}\otimes I_{\mathrm{B}}$ is added to $H_{\mathrm{unc}}$, the interaction
picture (\textquotedblleft toggling-frame\textquotedblright ) Hamiltonian 
$H_{\mathrm{tog}}(t)=U_{\mathrm{DD}}^{\dag }(t)H_{\mathrm{unc}}U_{\mathrm{DD}
}(t)$ reads 
\begin{equation}
H_{\mathrm{tog}}(t)=I_{\mathrm{S}}\otimes B_0 + f(t) \sigma_{z}
\otimes B_z,  
\label{hamil2}
\end{equation}
where the unitary $U_{\mathrm{DD}}(t)$ alternates between $I_{\mathrm{S}
}\otimes I_{\mathrm{B}}$ and $\sigma _{\bot }\otimes I_{\mathrm{B}}$ at the
instants $\{t_{j}\}_{j=1}^{N}$, and consequently the \textquotedblleft
switching function\textquotedblright\ $f(t)=\pm 1$ changes sign at the same
instants. 

We shall also need the magnitudes of the two parts of the Hamiltonian 
\begin{equation}
\label{Jbounds}
J_0 :=\Vert B_0\Vert <\infty ,\quad J_z:=\Vert B_z \Vert <\infty ,
\end{equation}
where $\Vert \cdot \Vert $ is the sup-operator norm (see Appendix 
\ref{app:norms}). There are certainly situations where either $J_0$ or $J_z$
can be divergent (e.g., $J_0$ in the case of oscillator baths). 
In such cases our bounds will not apply, and other methods such as 
correlation function bounds are more appropriate (see, e.g., Ref.\ 
\cite{NLP:09} for such bounds applied to DD).

\section{Time evolution bounds}

\label{sec:t-bound}

We aim to bound certain parts of the time evolution
operator induced by $H_\mathrm{tog}(t)$
\begin{equation}
\label{Uintegral}
U(T) = {\cal T}\exp\left( -i\int_0^T H_\mathrm{tog}(t)dt \right)
\end{equation}
where ${\cal T}$ is the time-ordering operator.

Standard time dependent perturbation theory provides
the following Dyson series for $U(T)$
\begin{subequations}
\label{Useries}
\begin{eqnarray}
\label{vec-sum}
U(T) &=& \sum_{n=0}^\infty (-iT)^n 
\sum_{\{\vec{\alpha};\mathrm{dim}(\vec{\alpha})=n\}} 
F_{\vec{\alpha}} \; \widehat Q_{\vec{\alpha}}
\\
\label{f_def}
F_{\vec{\alpha}} &:=& 
\int_0^1 ds_n f_{\alpha_n}(s_n) \int_0^{s_n} ds_{n-1} f_{\alpha_n}(s_n)\ldots 
\nonumber\\ 
&&\int_0^{s_3} ds_2  f_{\alpha_2}(s_{2}) \int_0^{s_2} ds_1 f_{\alpha_1}(s_{1})
\\
\widehat 
Q_{\vec{\alpha}} &:=& \sigma_{\alpha_n} B_{\alpha_n}   \ldots 
\sigma_{\alpha_2}B_{\alpha_2}   \sigma_{\alpha_1} B_{\alpha_1},
\label{q_def}
\end{eqnarray}
\end{subequations}
where $\mathrm{dim}(\vec{\alpha})=n$ is
the dimension of the vector $\vec{\alpha}$.
The identity $I_{\rm S}$
in the Hilbert space of the qubit/spin is denoted by $\sigma_0$.
In all sums over 
the vectors $\vec{\alpha}$ their components $\alpha_j$
take the values $0$ or $z$. In this way, the summation includes
all possible sequences of $B_0$ and $B_z$.
The function $f_0(s)$ is constant and equal to $1$ while $f_z(s):=f(s T)$
takes the valus $\pm1$. We use the dimensionless relative time
$s:=t/T$ so that all dependence on $T$ appears as a power in the prefactor.
Note that the coefficients $F_{\vec{\alpha}}$ do not depend on $T$.

In order to find an upper bound on \emph{each} term 
$F_{\vec{\alpha}} \; \widehat Q_{\vec{\alpha}}$ separately we proceed
in two steps. First, we use $|f_\alpha|= 1$ to obtain
\begin{eqnarray}
|F_{\vec{\alpha}}| &\le&  
\int_0^1 ds_n  \int_0^{s_n} ds_{n-1} \ldots 
\int_0^{s_3} ds_2  \int_0^{s_2} ds_1
\notag
\\
&=& \frac{1}{n!}.
\end{eqnarray}
Second, we use Eq.~\eqref{Jbounds} and $\Vert\sigma_\alpha\Vert =1$ 
to arrive at
\begin{eqnarray}
\Vert \widehat Q_{\vec{\alpha}} \Vert &\le& \prod_{j=1}^n
J_{\alpha_j} = J_0^{n-k(\vec{\alpha})} J_z^{k(\vec{\alpha})},
\end{eqnarray}
where we used the submultiplicativity of the sup-operator norm
(see Appendix~\ref{app:norms}).
The number $k(\vec{\alpha})$ stands for the number
of times that the factor $J_z$ occurs. Standard combinatorics of
binomial coefficients tells us that 
the term $J_0^{n-k} J_z^{k}$ occurs $n!/(k! (n-k)!)$ times
in the sum over all the vectors $\vec{\alpha}$ of given 
dimensionality $n$ in \eqref{vec-sum}. 
Hence each term of the  time expansion of $U(T)$ is 
bounded by
\begin{equation}
\Big\Vert \sum_{\{\vec{\alpha};\mathrm{dim}(\vec{\alpha})=n\}} 
F_{\vec{\alpha}} \; \widehat Q_{\vec{\alpha}} \Big\Vert
\le
\sum_{k=0}^n \frac{1}{k!(n-k)!} J_0^{n-k}J_z^k .
\label{bound1}
\end{equation}
We therefore define the bounding series
\begin{subequations}
\begin{eqnarray}
\label{S_def0}
S(J_0,J_z) &:=& \sum_{n=0}^{\infty} T^n 
\sum_{k=0}^n \frac{1}{k!(n-k)!} J_0^{n-k}J_z^k
\\
&=& \exp((J_0+J_z)T).
\label{S_def}
\end{eqnarray}
\end{subequations}
It then follows from Eq.~(\ref{bound1})
that each multinomial in $J_0$ and $J_z$ of the expansion of 
$S(J_0,J_z)$ is an upper bound on the norm of the sum of the corresponding
multinomial in the operators $B_0$ and $B_z$ of the expansion of $U(T)$
in Eq.~(\ref{vec-sum}). This is the property which we will use in the sequel.

\section{Bounds for dephasing}

\label{sec:dep-bound}

From  $\sigma_z^2=I_{\rm S}$ it is obvious that only the odd powers in $B_z$
contribute to dephasing while the even ones do not. Hence we split 
$U(T)$ as 
\begin{equation}
U(T)=I_{\rm S}\otimes B_{+}(T)+\sigma _{z}\otimes B_{-}(T)  
\label{Usplit}
\end{equation}
where the operators $B_{\pm }$ act only on the bath while $I_{\rm S}$ and 
$\sigma_{z}$ act only on the qubit. The operator $B_+$ comprises
all the terms with an even number $k$ of $\sigma_z \otimes B_z$, i.e., with
an even number of $J_z$ in the bounding series $S(J_0,J_z)$.  
The operator $B_-$ comprises
all the terms with odd number $k$ of $\sigma_z\otimes B_z$, i.e., with an odd
number of $J_z$ in the bounding series $S(J_0,J_z)$. 
Hence to bound the time series of $B_-(T)$ term by term we need the
the time series of the odd part of $S(J_0,J_z)$ in $J_z$. This, from
\eqref{S_def} is:
\begin{equation}
\label{Sminus_def}
S_-(J_0,J_z) = \exp(J_0 T)
\sinh(J_z T).
\end{equation}
The time series of $S_-(J_0,J_z)$ provides  a
bounding series of $B_-(T)$ term by term.
Hence we define
\begin{equation}
d_{k} := \frac{1}{k!} 
\frac{\partial ^{k}}{\partial T^{k}}S_{-}(J_0,J_z)\Big|_{T=0},
\end{equation}
such that $S_-(J_0,J_z)=\sum_{k=0} ^\infty d_k T^k$.

We know from the proof of Yang and Liu
\cite{Yang:08} that in the $B_0$-interaction picture 
a UDD sequence with $N$ pulses (which we denote by UDD($N$))
should make the first $N$ powers in $T$ of $B_-(T)$ vanish, i.e.,
$B_-(T)=O(T^{N+1})$.
However, since the Yang-Liu proof does not directly apply to our
discussion, we provide
a complete version of this proof which avoids the $B_0$-interaction picture
in Appendix~\ref{app:YL-proof}. The remaining powers
are bounded by the corresponding coefficients
$d_k$ of $S_-$. Thus the expression
\begin{equation}
\label{Delta_def}
\Delta_{N} := \sum_{k=N+1}^{\infty }d_{k}T^{k}
\end{equation}
provides an upper bound for $B_-(T)$ if UDD($N$) is applied:
\begin{equation}
\label{Bbound}
\Vert B_-(T) \Vert \le \Delta_N.
\end{equation}
Due to the obvious analyticity in the variable $T$
of $S_-(J_0,J_z)$ as defined in \eqref{Sminus_def}
we know that the residual term vanishes for
$N\to\infty$, i.e.,
\begin{equation}
\lim_{N\to \infty} \Delta_N = 0.
\end{equation}
This statement holds true irrespectively
of the values of $J_0$ and $J_z$, as long as they are finite.

We can obtain a more explicit expression for $\Delta _{N}$. Besides the
dimensionless number of pulses $N$ the bound $\Delta _{N}$ depends on 
$J_{0}T$ and on $J_{z}T$. It is convenient to introduce the dimensionless
parameters 
\begin{equation}
\varepsilon :=J_{0}T,\quad \eta := J_{z}/J_{0} 
\end{equation}
instead. In terms of these parameters we have
\begin{equation}
S_{-}(\eta ,\varepsilon )=\exp (\varepsilon )\sinh (\varepsilon \eta ).
\end{equation}
From the series
\begin{subequations}
\begin{eqnarray}
\exp (\varepsilon )\sinh (\varepsilon \eta ) &=&
\frac{1}{2}[e^{\varepsilon(1+\eta)}-e^{\varepsilon(1-\eta)}]
\\
&=& \sum_{l=0}^\infty \frac{\varepsilon^l}{2l!}[(1+\eta)^l-(1-\eta)^l]
\qquad
\\
&=& \sum_{l=0}^\infty p_l(\eta) \varepsilon^l
\end{eqnarray}
\end{subequations}
with
\begin{equation}
p_{l}(\eta ) := \frac{1}{2 l!}\left[(1+\eta)^l - (1-\eta)^l\right].
\label{p_l}
\end{equation}
we obtain
\begin{subequations}
\begin{eqnarray}
\label{Del-simp}
\Delta _{N}(\eta ,\varepsilon ) &=& \sum_{n=N+1}^{\infty}
p_n(\eta)\varepsilon^{n} 
\\
\label{Del-series}
&=& p_{N+1}(\eta )\varepsilon ^{N+1}+\mathcal{O}(\varepsilon ^{N+2}).
\end{eqnarray}
\end{subequations}
This, together with the bound \eqref{Bbound}, is our key result: 
it captures how the ``error'' $\Vert B_{-}(T)\Vert$ is suppressed 
as a function of the relevant
dimensionless parameters of the problem, $\eta $, $\varepsilon$, and $N$.
Note  that convergence for $N\to\infty$ is always ensured
by the factorial in the denominator, irrespectively
of the values of $\varepsilon$ and $\eta$ as long as these
are finite.

\begin{figure}[th]
\begin{center}
\includegraphics[width=\columnwidth]{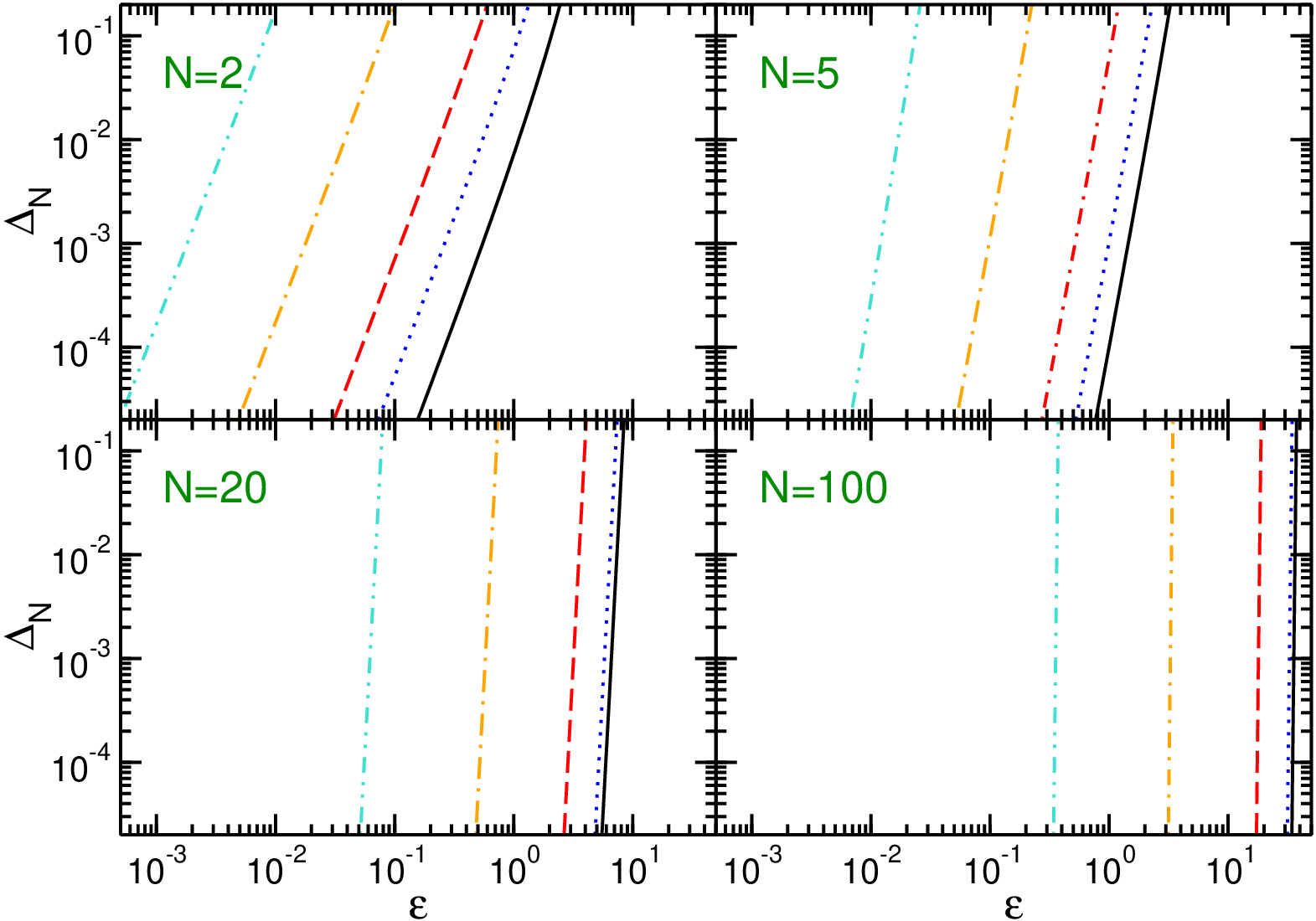}
\end{center}
\caption{(Color online) The bounding function $\Delta _{N}$
as a function of $\varepsilon =J_{0}T$, as given in
Eq.~\eqref{Delta_def} for various numbers of pulses $N$ and various 
values of the parameter $\protect\eta =J_z/J_0\in \{0.01,0.1,1,10,100\}$, 
with $\eta$ increasing from the rightmost curve to the leftmost curve in 
each panel.}
\label{fig1}
\end{figure}

\begin{figure}[th]
\begin{center}
\includegraphics[width=3.4in]{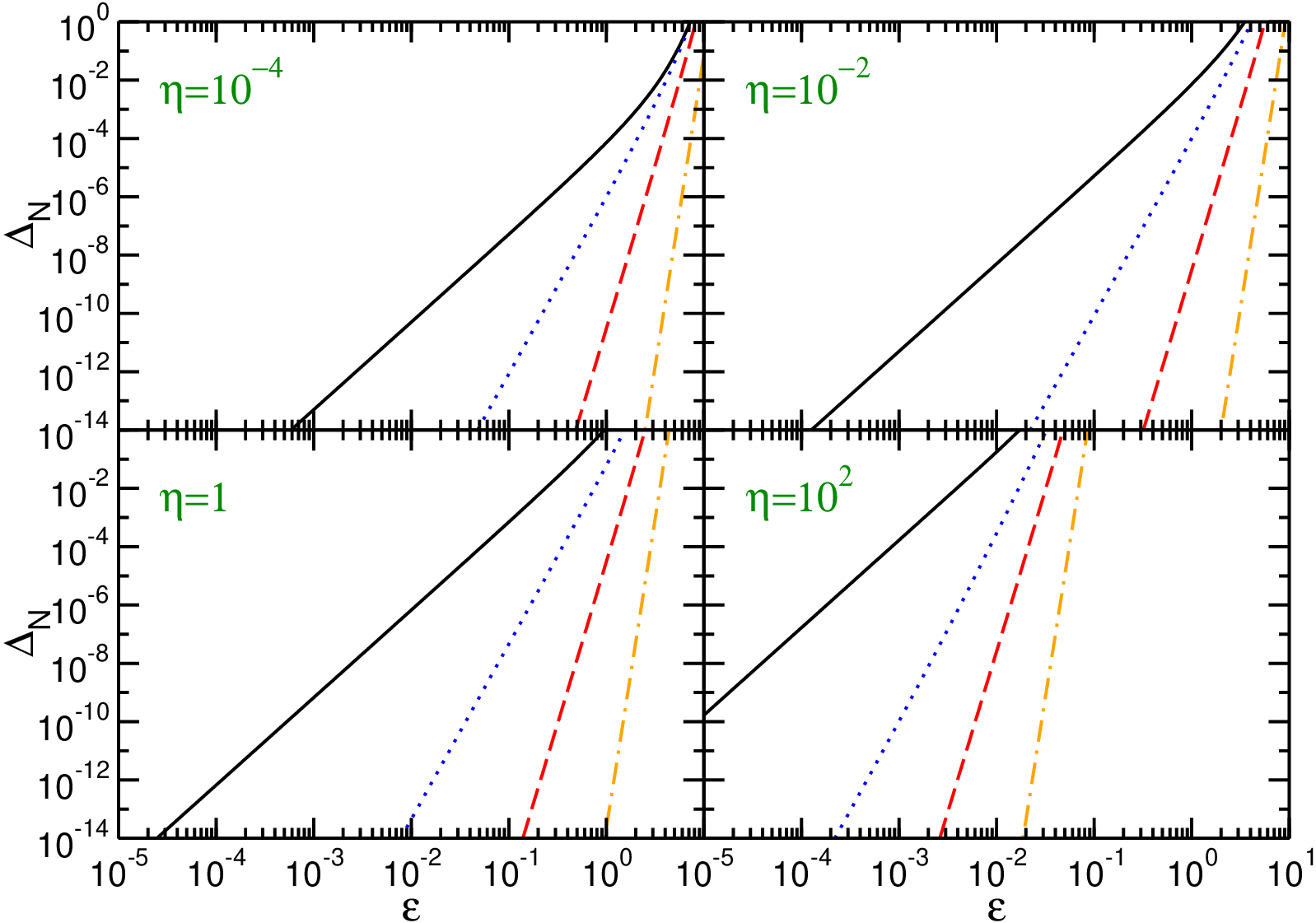}
\end{center}
\caption{(Color online) The bounding function $\Delta _{N}$
as a function of $\varepsilon =J_{0}T$, as given in Eq.~\eqref{Del-simp} for 
various numbers of pulses $N\in \{2,5,10,20\}$, at fixed values of 
$\protect\eta$. In each panel the curves become steeper as $N$ increases.
}
\label{fig2}
\end{figure}

For practical purposes it is advantageous not to compute $\Delta _{N}$
by the infinite series in \eqref{Del-simp}, but by 
\begin{equation}
\Delta _{N}(\eta ,\varepsilon )=S_{-}(\eta ,\varepsilon
)-\sum_{n=0}^{N} p_n(\eta)\varepsilon ^{n} ,  
\label{Delta_pract}
\end{equation}
which can easily be computed by computer
algebra programs. Figures~\ref{fig1} and \ref{fig2} depict the results of
this computation. Consider first Fig.~\ref{fig1}. Each curve shows
$\Delta _{N}(\eta ,\varepsilon )$ as a function of $\varepsilon$, at
fixed $\eta$ and $N$. The error $\Vert B_{-}(T)\Vert$ always
lies under the corresponding curve. Clearly, the bound becomes tighter as
$\varepsilon $ decreases. Moreover, the more pulses are
applied (the different panels) the higher the power in $\varepsilon$ and 
thus the steeper the curve. Additionally, the curves are shifted to the 
right as $N$ increases. Clearly, then, a larger number of pulses improves 
the error bound significantly, at fixed $\varepsilon$ and $\eta$.
This effect is even more conspicuous in Fig.~\ref{fig2}, where $\eta $ is 
fixed in each of the two panels, and the different curves correspond to 
different values of 
$N$. The vertical line intersects the bounding function at progressively
lower points as $N$ is increased, showing how the bound becomes tighter.

\section{Distance bound}

\label{sec:D-bound}

Intuitively, we expect the bound on $\|B_{-}(T)\|$ derived in the previous
section to be sufficient to bound the effect of dephasing. However, to make
this rigorous we need a bound on the trace-norm distance $D[\rho _{\mathrm{S}
}(T),\rho _{\mathrm{S}}^{0}(T)]$ between the \textquotedblleft
actual qubit\textquotedblright\ state 
\begin{equation}
\rho _{\mathrm{S}}(T) := \mathrm{tr}_{\mathrm B}[\rho_{\mathrm SB}(T) ]
\end{equation}
 and the
\textquotedblleft ideal qubit\textquotedblright\ state 
\begin{equation}
\rho _{\mathrm{S}}^{0}(T) := \mathrm{tr}_{\mathrm B}[\rho_{\mathrm SB}^0(T) ]
\end{equation}
where $\rho_{\mathrm SB}^0(T)$ is the time-evolved state without
coupling between qubit and bath.
The partial trace over the bath degrees of freedom is
a map from the joint system-bath Hilbert space to the 
system-only
Hilbert space (see Appendix~\ref{app:norms}), and is
denoted by ${\rm tr}_{\mathrm B}$.
As we shall see, the term $I_{\rm S}\otimes B_{+}(T)$ 
in Eq.~(\ref{Usplit}) indeed has a small, and in fact essentially
negligible effect.

To obtain the desired distance bound we consider a factorized initial state 
$\rho _{\mathrm{SB}}^{0}(0)=|\psi \rangle \langle \psi |\otimes \rho _{
\mathrm{B}}$, which evolves 
to $\rho _{\mathrm{SB}}(T)={U}(T)\rho _{\mathrm{SB}}^{0}(0){U}^{\dag }(T)$ 
when the system-bath interaction is on (the
\textquotedblleft actual\textquotedblright\ state), 
or to $\rho _{\mathrm{SB}}^{0}(T)=
I_{\rm S}\otimes U_{\mathrm B}(T)
\rho _{\mathrm{SB}}^{0}(0)
I_{\rm S}\otimes U_{\mathrm B}^\dag(T)$ 
when the interaction is off
(the \textquotedblleft ideal\textquotedblright\ state).
The unitary time evolution operator without coupling reads
\begin{equation}
U_{\mathrm B}(T):=\exp(-i T B_0),
\end{equation}
where $B_0$ is the pure-bath term in Eq.~(\ref{hamil1}).
The initial bath state 
$\rho _{\mathrm{B}}$ is arbitrary (e.g., a mixed thermal equilibrium state),
while the initial system state is pure. Let us define the correlation functions
\begin{equation}
b_{\alpha \beta }(T):=\mathrm{tr}
\left[ B_{\alpha }(T)\rho _{\mathrm{B}} B_{\beta }^{\dag }(T)\right]
\label{b_ab}
\end{equation}
where $\alpha,\beta\in\{+,-\}$,
and where all operators under the trace act only on the bath Hilbert space.
Explicit computation (see Appendix \ref{app:dist}) then yields: 
\begin{align}
& D[\rho _\mathrm{S}(T),\rho _\mathrm{S}^{0}(T)]  \label{D-bound-final} \\
& \leq \frac{1}{2}(|b_{++}(T)-1|+|b_{+-}(T)|+|b_{-+}(T)|+|b_{--}(T)|).\notag
\end{align}
We will show that $b_{++}$ is very close to $1$ while the other $b_{\alpha
\beta }$ quantities are small in the sense that they are bounded by
Eq.~\eqref{Bbound}.

First note from the unitarity of Eq.~\eqref{Usplit} that
\begin{eqnarray}
I &=& U^{\dagger }U 
\\
&=& I_{\rm S}\otimes (B_{+}^{\dagger }B_{+}+B_{-}^{\dagger }B_{-})
+\sigma _{z}\otimes (B_{-}^{\dagger }B_{+}+B_{+}^{\dagger }B_{-})\notag
\end{eqnarray}
where we omitted the time dependence $T$ to lighten the notation. Hence we
have 
\begin{subequations}
\begin{eqnarray}  \label{Bunitary2}
I &=&B_{+}^{\dagger }B_{+}+B_{-}^{\dagger }B_{-}  \label{Bunitary21} \\
0 &=&B_{+}^{\dagger }B_{-}+B_{-}^{\dagger }B_{+}.  \label{Bunitary22}
\end{eqnarray}
\end{subequations}
It follows that $\langle i|B_{+}^{\dagger }B_{+}|i\rangle =\Vert
B_{+}|i\rangle \Vert ^{2}\leq 1$ for all normalized states $|i\rangle $,
because $\langle i|B_{-}B_{-}^{\dagger }|i\rangle =\Vert B_{-}|i\rangle
\Vert ^{2}$ is non-negative. Thus in particular $\max_{\Vert |i\rangle \Vert
=1}\Vert B_{+}|i\rangle \Vert \leq 1$, and we can conclude that 
\begin{equation}
  \Vert B_{+}\Vert \leq 1.
  \label{Bunitary2a}
\end{equation}
Cyclic invariance of the trace in $b_{\alpha
\beta }$ together with Eq.~\eqref{Bunitary21} and the normalization 
$\mathrm{tr} [\rho _{\mathrm{B}}]=1$ 
immediately yields $b_{++}+b_{--}=1$, 
while the combination with Eq.~\eqref{Bunitary22} implies $b_{+-}+b_{-+}=0$. 
Hence Eq.~\eqref{D-bound-final} can be simplified to 
\begin{equation}
D[\rho _\mathrm{S}(T),\rho_\mathrm{S}^{0}(T)]\leq |b_{+-}(T)|+|b_{--}(T)|.
\label{D}
\end{equation}

To obtain a bound on the correlation functions $b_{\alpha \beta }$ we
use the following general correlation function inequality (for a proof
see Appendix~\ref{app:cor-ineq}):
\begin{eqnarray}
\left| \mathrm{tr}
\left[ Q\rho_\mathrm{B}Q^{\prime}\right]\right| \le\|Q^{\prime}\|\|Q\|,
\label{Q}
\end{eqnarray}
which holds for arbitrary bounded 
bath operators $Q,Q'$.
Applying Eq.~(\ref{Q}) to Eq.~\eqref{b_ab} yields
\begin{subequations}
\begin{eqnarray}
|b_{--}(T)| &\leq &\Vert B_{-}(T)\Vert ^{2}, \\
|b_{+-}(T)| &\leq &\Vert B_{+}(T)\Vert \Vert B_{-}(T)\Vert\\
& \leq &\Vert B_{-}(T)\Vert ,
\end{eqnarray}
\end{subequations}
where in the last inequality we used Eq.~\eqref{Bunitary2a}.

Summarizing, together with Eqs.~\eqref{Bbound} and \eqref{D} 
we have obtained the following
rigorous upper bound for the trace-norm distance
\begin{equation}
D[\rho _\mathrm{S}(T),\rho _\mathrm{S}^{0}(T)]\leq \min 
[1,\Delta _{N}(\eta ,\varepsilon)+\Delta _{N}^{2}(\eta ,\varepsilon )].  
\label{Dbound}
\end{equation}
This upper bound completes our main result. Since as we saw
in Eq.~\eqref{Del-series} $\Delta_{N}(\eta ,\varepsilon )=
p_{N+1} (\eta )\varepsilon ^{N+1}+\mathcal{O} (\varepsilon ^{N+2})$, the
appearance of the squared term in Eq.~\eqref{Dbound} 
(whose origin is $|b_{--}(T)|$) is not relevant in the sense that even in
the presence of this term the bound
\begin{equation}
D[\rho _\mathrm{S}(T),\rho _\mathrm{S}^{0}(T)]\leq p_{N+1}
(\eta )\varepsilon ^{N+1}+\mathcal{O}(\varepsilon ^{N+2})
\end{equation}
holds. Hence the result of Eq.~\eqref{Del-simp}
depicted in Figs.\ \ref{fig1} and \ref{fig2} provides the desired
result. Ignoring the $\Delta _{N}^{2}$ term 
in Eq.~(\ref{Dbound}), we note that Figs.\ \ref{fig1} and \ref{fig2} also
reveal the limitations of our bound when $\varepsilon $ or $\eta $ are 
too large for a given value of $N$:
For any pair of states it is always the case
that $D\leq 1$, so that as soon as $\Delta _{N}=1$ the bound no longer
provides any useful information.

Note further that the results shown in Fig.\ \ref{fig1} are qualitatively
similar to the results obtained for the analytically solvable
spin-boson model for pure dephasing \cite{Uhrig:07}.
Heuristically, the necessary
identification is $J_0 =\omega_{\mathrm{D}}$ where $\omega_{\mathrm{D}}$
is the hard cutoff of the spectral function and $\eta \propto \alpha $
where $\alpha $ is the dimensionless coupling constant for Ohmic noise. We
stress that the advantage of Eq.~\eqref{Dbound} compared to the analytically
exact results in Ref.\ \cite{Uhrig:07} is that it holds rigorously for a
large class of pure dephasing models, namely those of bounded Hamiltonians.

\section{Analysis for finite minimum pulse interval}

\label{analysis}

So far we have essentially treated the total time $T$ and the number of
pulses $N$ as independent parameters. This is possible when there is no
lower limit on the pulse intervals. However, in reality this is never the
case and in this section we analyze what happens when there is such a lower
limit. Note that it follows from Eq.~(\ref{udd}) that the smallest pulse
interval is the first: $t_{1}=T\sin ^{2}(\pi /(2N+2))$. Let us assume that 
$t_{1}$ is fixed, so that, given $t_{1}$ and $N$, the total time is
\begin{subequations}
\begin{eqnarray}
T(N) &=&t_{1}q(N), \\
q(N) &:=&\csc ^{2}\left( \frac{\pi }{2N+2}\right) .
\end{eqnarray}
\end{subequations}

For large $N$ we can expand the $\csc ^{2}$ function to first order in its
small argument, yielding 
\begin{equation}
q(N)=\left( \frac{2N+2}{\pi }\right) ^{2}+
\frac{1}{3}+{\cal O}\left( N^{-2}\right) ,
\end{equation}
which shows how the total time grows as a function of $N$ at fixed minimum
pulse interval $t_{1}$. Along with $\eta $, the relevant dimensionless
parameter is now
\begin{equation}
\varepsilon_{1}:=J_0 t_{1},
\end{equation}
instead of $\varepsilon = q(N)\varepsilon _{1}$.
We can then rewrite the bounding function \eqref{Del-simp} 
in terms of these quantities as
\begin{equation}
\label{Del1}
\Delta _{N}(\eta ,\varepsilon _{1}) =
\sum_{n=N+1}^\infty p_n(\eta)q^n(n)\varepsilon_1^n .
\end{equation}
Considering now the large $N$ limit of the first term in this sum, we have
\begin{eqnarray}
p_{N+1}(\eta)q^{N+1}(N+1)\varepsilon_1^{N+1} &\approx &
\frac{1}{2N!}(\frac{2}{\pi}N)^{2N}[(1+\eta)\varepsilon_1]^N \notag \\
\label{Del1-1}
&\approx & (cN)^N
\end{eqnarray}
where we kept only the leading order terms and neglected all additive 
constants relative
to $N$, and in Eq.~\eqref{Del1-1} used Stirling's approximation
$n! \approx (n/e)^n$. The constant $c$ is
$\frac{1}{2}(\frac{2}{\pi})^2 e (1+\eta)\varepsilon_1$.
We thus see clearly that for fixed $t_{1}$ it becomes
counterproductive to make $N$ too large, since
no matter how small $c$ is, for large enough $N$ the factor $N^N$ will 
eventually dominate.
This reflects the competition between the gains due to higher order
pulse sequences and the losses due to 
the increased coupling time to the qubit allotted to the environment.
Similar conclusions, delineating regimes where increasingly long DD
sequences become disadvantageous, have been reported for periodic 
\cite{KhodjastehLidar:07,KhodjastehLidar:08} and concatenated 
\cite{KhodjastehLidar:07,NLP:09,West:09} DD pulse sequences, as well as for the
QDD sequence \cite{WFL:09}.

\begin{figure}[th]
\begin{center}
\includegraphics[width=3.4in]{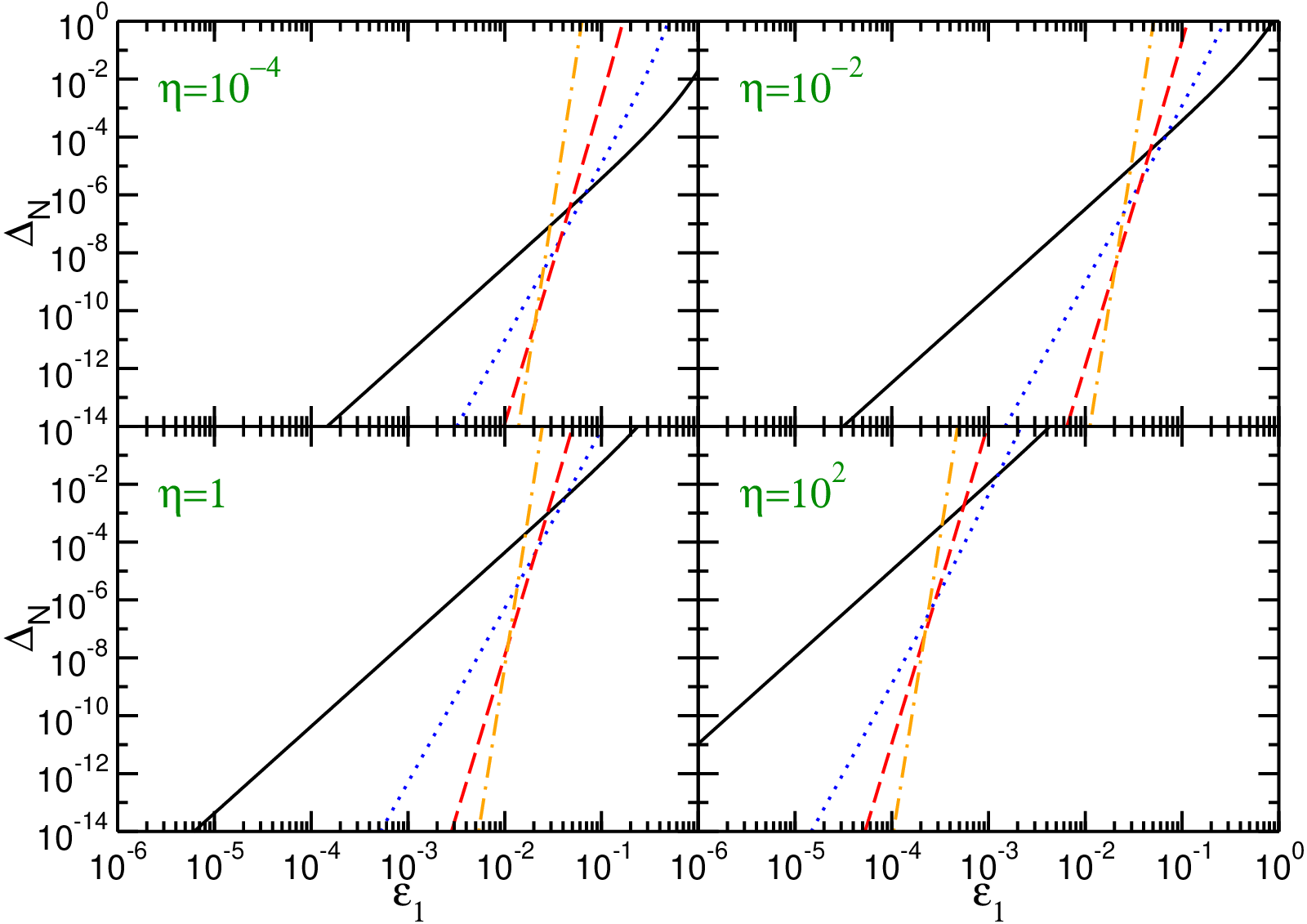}
\end{center}
\caption{
(Color online) The bounding function $\Delta _{N}$ 
as a function of $\varepsilon_1=t_1 J_0$ (where $t_1$ is the smallest
pulse interval), at fixed values of $\eta$.
The number of pulses $N\in \{2,5,10,20\}$ is varied from curve 
to curve. In each panel the curves become steeper as $N$ increases.
}\label{fig3}
\end{figure}

These conclusions are further illustrated in Fig.~\ref{fig3}, where we plot
the bound $\Delta _{N}(\eta ,\varepsilon _{1})$ by replacing $\varepsilon $
with $\varepsilon _{1}q(N)$ in Eq.~\eqref{Del-simp}. This figure should be
contrasted with Fig.~\ref{fig2}. The most notable change is that increasing 
$N$ now no longer uniformly improves performance. Whereas in Fig.~\ref{fig2}
the curves for different values of $N$ all tend to converge at high values
of $\varepsilon $, in Fig.~\ref{fig3} a high $N$ value results in a steeper
slope, but also moves the curve to the left. Thus, for a fixed value of 
$\varepsilon _{1}$ it can be advantageous to use a small value of $N$ (e.g.,
for $\eta =0.01$ and $\varepsilon _{1}=0.1$ the $N=2$ curve provides the
tightest bound).

\section{Conclusions}

\label{sec:conc}

We have derived rigorous performance bounds for the UDD sequence protecting
a qubit against pure dephasing. The derivation is based on
the existence of finite bounds for the relevant parts of the
Hamiltonian, captured in the dimensionless parameters $\varepsilon$ and $\eta$.
Under this assumption the bounds show rigorously that  
dephasing is suppressed to leading order as $(1/N!)[\varepsilon (1+\eta)]^{N}$
We consider it a vital step to know that irrespectively of any details of the
bath, except for the existence of finite bounds, a large number $N$ of
pulses is always advantageous  at fixed $T$ -- at least under the
idealized assumption of perfect and instantaneous pulses.

An immediate corollary of our results is that identical bounds apply
for the case of the UDD sequence protecting a qubit against
longitudinal relaxation. This is the case when the uncontrolled
Hamiltonian (\ref{hamil1}) is replaced by $H_{\mathrm{unc}}=I_{\mathrm{S}}
\otimes H_{\mathrm{B}}+\sigma _{\bot }\otimes B$, and the UDD pulse sequence
consists of rotations about the spin-$z$ axis.
A practical implication is that the bounds found here can be used
to check numerical and approximate calculations.
Such calculations must obey our mathematically rigorous bounds,
so that a testbed is provided.

Furthermore, a number of interesting
generalizations and extensions of our results readily suggest themselves.
One is to consider rigorous bounds for finite pulse-width UDD sequences. It
is already known how to construct such sequences with pulse-width errors
which appear only to third order in the value of the pulse width 
\cite{Uhrig:09a}, but no rigorous bounds have been found. Another important
generalization, as mentioned above, is to the QDD sequence for general
decoherence \cite{WFL:09,Pasini:09}. 
We expect that techniques similar to the ones we
introduced here will apply to both of these open problems. Yet another
direction, which will require different techniques, is to find rigorous UDD
performance bounds for unbounded baths, such as oscillator baths. It is
likely that a correlation function analysis similar to that performed in
Ref.~\cite{NLP:09} for periodic and concatenated DD sequences will prove
useful in this case.

\begin{acknowledgments}
G.S.U.  is supported under DFG grant UH 90/5-1. D.A.L. is supported
under NSF grants CHE-924318 and PHY-802678,
and by the United States Department of Defense. 
\end{acknowledgments}

\appendix{}

\section{Norms and Distances}

\label{app:norms}

\subsection{Trace and partial trace}
We deal only with linear trace-class bounded operators that map
between separable Hilbert spaces in this work. A Hilbert space ${\cal
H}$ is separable if and only if it admits a countable orthonormal
basis. A bounded linear operator $A:{\cal H}\mapsto {\cal H}$, where
${\cal H}$ is separable, is said to be in the trace class if for some
(and hence all) orthonormal bases $\{|k\rangle\}_k$ of ${\cal H}$ the
sum of positive terms $\sum_k \langle k|A^\dagger A|k\rangle$ is
finite. In this case, the sum $\sum_k \langle k|A|k\rangle$ is
absolutely convergent and is independent of the choice of the
orthonormal basis. This value is called the trace of $A$, denoted by
${\rm tr}(A)$. Whenever we use the symbol ${\rm tr}$ in this work, we
mean the trace over the full Hilbert space the operator the trace is
taken over is acting on.

Now consider two separable Hilbert spaces ${\cal H}_1$ and ${\cal
H}_2$ and let $A:{\cal H}\mapsto {\cal H}$ denote a linear trace-class
bounded operator acting on the tensor product Hilbert space ${\cal
H}:={\cal H}_1\otimes {\cal H}_2$. Let $\{|k_i\rangle\}_k$ denote an
orthonormal basis for ${\cal H}_i$, where $i\in{1,2}$. The partial
trace operation over the first (second) Hilbert space is a map from
${\cal H}$ to the second (first) Hilbert space, and has the
operational definition ${\rm tr}_i(A):=\sum_{k_i} \langle
k_i|A|k_i\rangle$. When $A$ is decomposed in terms of the two
orthonormal bases as $A = \sum_{k_1,k'_1,l_2,l'_2} \langle
k_1l_2|A|k'_1l'_2\rangle |k_1l_2\rangle\langle k'_1l'_2|$, where
$|k_1l_2\rangle := |k_1\rangle \otimes |l_2\rangle$ etc., the partial
trace over ${\cal H}_2$ can written as ${\rm
tr}_2(A)=\sum_{k_1,k'_1,l_2} \langle k_1l_2|A|k'_1l_2\rangle|k_1\rangle
\langle k'_1|$. This makes it clear that ${\rm tr}_2(A)$ is an
operator that acts on ${\cal H}_1$.

\subsection{Sup-operator norm and trace-norm}

We make frequent use of two matrix norms \cite{Bhatia:book} in this work.
The first is the sup-operator norm 
\begin{equation}
\Vert A\Vert _{\infty }:= \sup_{\left\Vert |v\rangle \right\Vert
=1}\left\Vert A|v\rangle \right\Vert =\sup_{|v\rangle }\sqrt{\langle
v|A^{\dag }A|v\rangle }/\sqrt{\langle v|v\rangle }
\end{equation}
The sup-operator norm of $A$ is the largest eigenvalue of $|A|:= 
\sqrt{A^{\dag }A}$, i.e., the largest singular value of $A$. Since we use 
it often  we denote $\Vert A\Vert _{\infty }$ for simplicity by 
$\Vert A\Vert $,  and context
should make it clear whether we are taking the norm of an operator or simply
the Euclidean norm $\left\Vert |v\rangle \right\Vert := \sqrt{\langle
v|v\rangle }$ of a vector $|v\rangle $. Note that if $A$ is normal 
($A^{\dag}A=AA^{\dag}$, it can be unitarily diagonalized, 
so that $A^{\dag }A=VD^{\dag}DV^{\dag }$ where $V$ is unitary and $D$ 
is the diagonal matrix of
eigenvalues of $A$), the largest singular value coincides with the largest
absolute value of the eigenvalues of $A$, i.e., $\Vert A\Vert
=\sup_{\left\Vert |v\rangle \right\Vert =1}|\langle v|A|v\rangle |$.

The trace-norm 
\begin{equation}
\Vert A\Vert _{1}:= \mathrm{tr}\sqrt{A^{\dag }A}
\end{equation}
is the sum of the eigenvalues of $|A|$, i.e., the sum of the singular values
of $A$. Therefore $\Vert A\Vert \leq \Vert A\Vert _{1}$. Both norms are
unitarily invariant ($\Vert VAW\Vert _{\mathrm{ui}}=
\Vert A\Vert _{\mathrm{ui}}$ 
for any pair of unitaries $V$ and $W$) and therefore submultiplicative 
($\Vert AB\Vert _{\mathrm{ui}}\leq \Vert A\Vert_{\mathrm{ui}}
\Vert B\Vert_{\mathrm{ui}}$) \cite{Bhatia:book}. 
In this work we make frequent use of both
properties. In addition unitarily invariant norms are invariant under
Hermitian conjugation, i.e., $\Vert A\Vert _{\mathrm{ui}}=\Vert A^{\dag
}\Vert _{\mathrm{ui}}$. This follows from the singular value decomposition:\ 
$A=V\Sigma W^\dag$, where $V$ and $W$ are unitaries and $\Sigma $ is the 
diagonal
matrix of singular values of $A$. Since the singular values are all positive
we have $A^{\dag }=W\Sigma V^{\dag }$ and hence $\Vert A^{\dag
}\Vert _{\mathrm{ui}}=\Vert \Sigma \Vert _{\mathrm{ui}}=
\Vert A\Vert_{\mathrm{ui}}$.

\subsection{Trace-norm distance and fidelity}

The trace-norm distance between two mixed states 
described by the density operators $\rho _{1}$ and $\rho _{2}$ is
defined as
\begin{equation}
D[\rho _{1},\rho _{2}]:= \frac{1}{2}\Vert \rho _{1}-\rho _{2}\Vert _{1}.
\end{equation}
It is bounded between $0$ and $1$, vanishes if and only
if $\rho _{1}=\rho _{2}$ and is $1$ if and only if 
$\rho _{1}$ are $\rho _{2}$ are orthogonal, i.e., ${\rm  tr}(\rho_1\rho_2)=0$. 

The trace-norm distance is a standard and useful measure
of distinguishability between states. The reason is this: Assume that we
perform a generalized measurement (POVM --
positive operator valued measurement) $E$ with corresponding measurement
operators $\{E_{i}\}$ satisfying the normalization condition 
$\sum_{i}E_{i}=I $. The measurement outcomes are described by the the
measurement probabilities $p_{i}=\mathrm{tr}[\rho _{1}E_{i}]$ and $q_{i}=
\mathrm{tr}[\rho _{2}E_{i}]$. The Kolmogorov distance between the two
probability distributions produced by these measurements is $K_{E}(p,q)
={\frac{1}{2}}\sum_{i}|p_{i}-q_{i}|$, and it can be shown that 
$D[\rho_{1},\rho _{2}]=\max_{E}K_{E}$, i.e., the trace-norm distance equals the
maximum over all possible generalized measurements of the Kolmogorov
distance between the probability distributions resulting from measuring 
$\rho _{1}$ and $\rho _{2} $ \cite{Nielsen:book}. The trace-norm distance is
related to the Uhlman fidelity 
\begin{equation}
F[\rho _{1},\rho _{2}]:= \Vert \sqrt{\rho _{1}}\sqrt{\rho _{2}}\Vert
_{1}=\mathrm{tr}\sqrt{\sqrt{\rho _{1}}\rho _{2}\sqrt{\rho _{1}}}
\end{equation}
via 
\begin{equation}
1-D\leq F\leq \sqrt{1-D^{2}},
\end{equation}
so that one bounds the other \cite{Fuchs:99}.

\section{Proof of the vanishing orders in UDD}

\label{app:YL-proof}

The paper by Yang and Liu \cite{Yang:08} sketches a proof of the
universality of UDD in the interaction picture. In this appendix we provide
the first comprehensive proof of the Yang-Liu universality result. Our proof
is done in the toggling frame rather than the interaction picture.

We shall prove that all powers $n\leq N$ vanish in the expansion of the time
evolution operator in \eqref{Useries} which have an \emph{odd} number of 
$\sigma _{z}B_{z}$ in $\widehat{Q}_{\vec{\alpha}}$, thus also an odd number
of $f_{z}$ in $F_{\vec{\alpha}}$. This is equivalent to showing that the
first $N$ powers in $T$ of $B_{-}(T)$ vanish, i.e., that dephasing occurs
only in order $T^{N+1}$ or higher. Henceforth we use the shorthand 
$\bar{N}:=N+1$.

The substitution $s =\sin^2(\theta/2)$ suggests itself based on the UDD
choice for the $\{ \delta_j \}$ in \eqref{udd}, because it renders 
\begin{equation}
\tilde f_\alpha(\theta):=f_\alpha(\sin^2(\theta/2))
\end{equation}
particularly simple if the $\{ \delta_j \}$ are chosen according to 
Eq.~
\eqref{udd}: 
\begin{equation}
\tilde{f}_z(\theta)=(-1)^j
\end{equation}
holds for $\theta\in (j\pi/\bar N,(j+1)\pi/\bar N)$ with $j\in\{0,\ldots,N\}$%
. For simplicity, we will omit the tilde on the functions $f_\alpha$ from
now on because only the argument $\theta$ will appear henceforth.

Since $f(\theta)$ enters the nested integrals only with an argument in $%
[0,\pi]$ it does not matter what we assume about $f(\theta)$ outside the
limits of these integral, and we release the constraint on $j$, allowing $%
j\in\mathbb{Z}$. The function $f_z(\theta)$ then becomes an odd function
with antiperiod $\pi/\bar N$. Thus its Fourier series 
\begin{equation}  
\label{fourier}
f_z(\theta) = \sum_{k=0}^\infty c_{2k+1} \sin((2k+1)\bar N\theta)
\end{equation}
contains only harmonics $\sin(r\bar N\theta)$ with $r$ an odd integer. The
precise coefficients $c_{2k+1}$ do not matter, a fact which can be exploited
for other purposes, e.g., to deal with pulses of finite duration 
\cite{Uhrig:09a}.

Under the substitution $s =\sin^2(\theta/2)$ the infinitesimal element $ds$
becomes $ds\to \frac{1}{2}\sin(\theta)d\theta$, converting \eqref{f_def}
into \begin{eqnarray}  \label{f_alpha}
&& F_{\vec{\alpha}} = \\
&& \int_0^\pi \sin(\theta_n) d\theta_n f_{\alpha_n}(\theta_n)
\int_0^{\theta_n} \sin(\theta_{n-1})d\theta_{n-1}
f_{\alpha_{n-1}}(\theta_{n-1})  \notag \\
&&\ldots \int_0^{\theta_3} \sin(\theta_2)d\theta_2 f_{\alpha_2}(\theta_{2})
\int_0^{\theta_2} \sin(\theta_1) d\theta_1 f_{\alpha_1}(\theta_1),  \notag
\end{eqnarray}
where we absorbed the $\frac{1}{2}$ factors coming from the infinitesimal
elements into the coefficients $c_{2k+1}$.

What happens if we perform the successive integrations in 
Eq.~\eqref{f_alpha}? 
Replacing $f_{z}(\theta )$ by its Fourier series \eqref{fourier} we deal
with integrands which are products of trigonometric functions. The
substitution gave rise to the factor $\sin \theta $. The Fourier series
gives rise to additional factors $\sin (r_{o}\bar{N}\theta )$, where $r_{o}$
is an odd integer. Recall the elementary trigonometric identities 
\begin{subequations}
\begin{eqnarray}
\sin a\sin b &=&\frac{1}{2}\left[ \cos \left( a-b\right) -\cos \left(
a+b\right) \right] ,  \label{sinsin} \\
\cos a\sin b &=&\frac{1}{2}\left[ \sin \left( a+b\right) -\sin \left(
a-b\right) \right] ,  \label{cossin} \\
\cos a\cos b &=&\frac{1}{2}\left[ \cos \left( a+b\right) +\cos \left(
a-b\right) \right] .  \label{coscos}
\end{eqnarray}
\end{subequations}
Using these, the most general trigonometric factor to occur in the course of
the integrations in Eq.~\eqref{f_alpha} can be written as either 
$\sin [(q+r\bar{N})\theta ]$ or $\cos [(q+r\bar{N})\theta ]$ where $r$ and 
$q$ are integers. Since we are only concerned with values of $n$ such that 
$n<\bar{N}$, the absolute value of $q$ always remains below $\bar{N}$, so 
that the representation of the integer factor $(q+r\bar{N})$ in the arguments 
of the trigonometric functions is unique.

We now consider a complete set of four different cases which can occur in
the course of the evaluation of each $F_{\vec{\alpha}}$. The first two cases
are associated with the occurrence of $f_{0}=1$ in one or more of the nested
integrals. Suppose for concreteness that this happens in nested integral
number $j$. Then the factor $\sin (r_{o}\bar{N}\theta _{j})$ does not occur
in this integral, since this factor arises exclusively due to the presence
of $f_{z}(\theta _{j})$. The two cases are now distinguished by whether a
summand in the integrand of this $j$th integral, after a complete expansion
of trigonometric products \emph{excluding the} $sin(\theta _{j})$ \emph{term}, 
into sums using Eqs.~\eqref{sinsin}-\eqref{coscos}, involves the factor 
$\cos [(q+r\bar{N})\theta ]$ (whence we call the integrand \textquotedblleft
cosine-type\textquotedblright ) or the factor $\sin [(q+r\bar{N})\theta ]$
(whence we call the integrand \textquotedblleft sine-type\textquotedblright
). The third and fourth cases are associated with the occurrence of $f_{z}$
in integrand number $j$. Then the factor $\sin (r_{o}\bar{N}\theta _{j})$
does occur in this integrand, and again we distinguish two cases according
to the presence of $\cos [(q+r\bar{N})\theta ]$ (\textquotedblleft
cosine-type\textquotedblright ) or $\sin [(q+r\bar{N})\theta ]$
(\textquotedblleft sine-type\textquotedblright ) arising from a complete
expansion of trigonometric products \emph{excluding the} $sin(\theta _{j})$ 
\emph{term and also the }$\sin (r_{o}\bar{N}\theta _{j})$\emph{\ term}.
Here, then, are the four cases in detail:

\begin{enumerate}
\item[(i)] Assume that one of the nested integrals contains $f_{0}=1$ and
the factor $\cos [(q+r\bar{N})\theta ]$. As we shall see in item 4) below
this case occurs when $r$ is even. Then this integral reads 

\begin{eqnarray}
&&2\int \cos [(q+r\bar{N})\theta ]\sin (\theta )d\theta =  \label{cos-ohne}
\\
&&\frac{\cos [(-1+q+r\bar{N})\theta ]}{-1+q+r\bar{N}}-\frac{\cos [(1+q+r
\bar{N})\theta ]}{1+q+r\bar{N}}  \notag
\end{eqnarray}
In writing Eq.~\eqref{cos-ohne} we have assumed that the denominators do not
vanish. The denominators may in fact vanish because $r$ may be zero. 
When $r=0$ the case $|q|=1$ is special and yields 
\begin{equation}
2\int \cos (\pm \theta )\sin (\theta )d\theta =-\frac{1}{2}\cos (2\theta ).
\label{cos-ohne2}
\end{equation}
The important point is that both Eq.~\eqref{cos-ohne} and \eqref{cos-ohne2}
have only cosine terms on the right hand side.

\item[(ii)] Assume that one of the nested integrals contains $f_{0}=1$ and
the factor $\sin [(q+r\bar{N})\theta ]$. As we shall see in item 4) below
this case occurs when $r$ is odd. Then this integral reads 
\begin{eqnarray}
&&2\int \sin [(q+r\bar{N})\theta ]\sin (\theta )d\theta =  \label{sin-ohne}
\\
&&\frac{\sin [(-1+q+r\bar{N})\theta ]}{-1+q+r\bar{N}}-\frac{\sin [(1+q+
r\bar{N})\theta ]}{1+q+r\bar{N}}.  \notag
\end{eqnarray}

No denominator can vanish because as we shall see in item 2) below, $|q|<N$.
The important point here is that Eq.~\eqref{sin-ohne} has only sine terms on
the right hand side.

\item[(iii)] Assume that one of the nested integrals contains $f_{z}$ and
the factor $\cos [(q+r\bar{N})\theta ]$. As we shall see in item 4) below
this case occurs when $r$ is even. Then this integral reads 
\begin{eqnarray}
&&4\int \cos [(q+r\bar{N})\theta ]\sin (r_{o}\bar{N}\theta )\sin (\theta
)d\theta =  \notag  \label{cos-mit} \\
&&\qquad \frac{\sin [(-1+q+(r+r_{o})\bar{N})\theta ]}{-1+q+(r+r_{o})\bar{N}}
\notag \\
&&\qquad -\frac{\sin [(1+q+(r+r_{o})\bar{N})\theta ]}{1+q+(r+r_{o})\bar{N}} 
\notag \\
&&\qquad -\frac{\sin [(-1+q+(r-r_{o})\bar{N})\theta ]}{-1+q+(r-r_{o})\bar{N}}
\notag \\
&&\qquad +\frac{\sin [(1+q+(r-r_{o})\bar{N})\theta ]}{1+q+(r-r_{o})\bar{N}}.
\end{eqnarray}
Since $r\pm r_{o}$ is odd none of the denominators can vanish as long as 
$|q|<N$. Again, the important point here is that Eq.~\eqref{cos-mit} has only
sine terms on the right hand side.

\item[(iv)] Finally, assume that one of the nested integrals contains $f_{z}$
and the factor $\sin [(q+r\bar{N})\theta ]$. As we shall see in item 4)
below this case occurs when $r$ is odd. Then this integral reads 
\begin{eqnarray}
&&4\int \sin [(q+r\bar{N})\theta ]\sin (r_{o}\bar{N}\theta )\sin (\theta
)d\theta =  \notag  \label{sin-mit} \\
&&\qquad \frac{\cos [(-1+q+(r-r_{o})\bar{N})\theta ]}{-1+q+(r-r_{o})\bar{N}}
\notag \\
&&\qquad -\frac{\cos [(1+q+(r-r_{o})\bar{N})\theta ]}{1+q+(r-r_{o})\bar{N}} 
\notag \\
&&\qquad -\frac{\cos [(-1+q+(r+r_{o})\bar{N})\theta ]}{-1+q+(r+r_{o})\bar{N}}
\notag \\
&&\qquad +\frac{\cos [(1+q+(r+r_{o})\bar{N})\theta ]}{1+q+(r+r_{o})\bar{N}}.
\end{eqnarray}
In writing Eq.~\eqref{sin-mit} we have assumed that the denominators do not
vanish. A denominator can vanish only when $|q|=1$, which leads to the two
special cases $r=\pm r_{o}$. In analyzing these two cases we can assume
without loss of generality that $q=1$
and $r=r_{o}$. Otherwise we multiply the argument of the first
and/or the second sine-function in the integrand by $-1$.
This yields
\begin{eqnarray}
&&4\int \sin [(1+r\bar{N})\theta ]\sin (r\bar{N}\theta )\sin (\theta
)d\theta =  \label{sin-mit2} \\
&&\quad -\frac{\cos (2\theta )}{2}-\frac{\cos (2r\bar{N}\theta )}{2r\bar{N}}+
\frac{\cos [(2+2r\bar{N})\theta ]}{2+2r\bar{N}}.  \notag
\end{eqnarray}
The important point here is that in 
Eq.~\eqref{sin-mit}
only cosine terms appear on the right hand side.
\end{enumerate}

The number of possible terms proliferates in the course of the successive
integrations. Therefore, in the sequel we discuss only the common features
of the resulting summands. It is always understood that sums with varying
sets of $q$ and $r$ are considered. We present a series of observations
which leads to the desired proof of the cancellation of the first $N$ powers
in $T$ of $B_{-}(T)$. The key to doing so will be to show that after
integrating with an odd number of $f_{z}$ factors we always end up with a
sine-type integrand. 

\begin{enumerate}
\item[1)] Recall that we call an integrand summand \textquotedblleft
cosine-type\textquotedblright\ or \textquotedblleft
sine-type\textquotedblright\ if after complete expansion of all
trigonometric products, excluding the $\sin(\theta )$ term arising from the
change of variables and of the $f_{z}$ term if it is there, the
trigonometric factor is $\cos [(q+r\bar{N})\theta ]$ or 
$\sin [(q+r\bar{N})\theta ]$, respectively. 
Cases (i) and (iii) above are cosine-type, while
cases (ii) and (iv) are sine-type. It is clear that the first integrand in 
\eqref{f_alpha} is cosine-type with $r=0$ and $q=0$.

\item[2)] We track which values of $q$ may occur in each integration. The
first integrand in \eqref{f_alpha} is either $\sin (\theta _{1})$ or $\sin
(\theta _{1})\sin (r\bar{N}\theta _{1})$, i.e., it has $q=0$, so that this
is our starting point. It follows from Eqs.~\eqref{cos-ohne}-\eqref{sin-mit2}
that each integration increments the possible maximum of $|q|$ by unity. The
highest power in $T$ studied is $T^{N}$ so that there are $n\leq N$
integrations. This implies that the final value $q_{\mathrm{final}}$ before
the very last integration obeys $|q_{\mathrm{final}}|<N$.

\item[3)] We track whether even or odd values of $r$ occur at each
integration. As mentioned in item 1), $r=0$ holds in the first integration,
so that our starting point is an even value. Each integration involving $%
f_{0}$ leaves $r$ unchanged. Each integration involving $f_{z}$ adds or
substracts $r_{o}$, so that $r$ changes from even to odd or vice versa. If
we combine this with the
results of cases (i)-(iv) [Eqs.~\eqref{cos-ohne}-\eqref{sin-mit2}
this reveals the input-output table in Eq. (\ref{tab}).
The integrands (i.e., the inputs) are indicated by the case number in the
table entries, while the values of the integrals (i.e., the outputs) are the
types indicated in corresponding table entries. Also indicated is the
transformation undergone by $r$ from input to output.
\begin{equation}
\begin{tabular}{l||l|l}
$\times $ & cosine-type & sine-type \\ \hline\hline
$f_{0}$ & case (i): & case (ii):\  \\ 
& cosine-type,\ $r\mapsto r$ & sine-type,\ $r\mapsto r$ \\ \hline
$f_{z}$ & case (iii):\  & case (iv):  \\ 
& sine-type,\ $r\mapsto r\pm r_{o}$ & cosine-type,\ $r\mapsto r\pm
r_{o}$
\end{tabular}
\label{tab}
\end{equation}

\item[4)] Consider the output of the table as the input into the next
integration and focus on the $f_{z}$ row. Note that cases (iii) and (iv)
alternate along with a change in parity of $r$, i.e., cosine-type 
changes into sine-type 
and vice versa, while odd $r$ changes
to even $r$ and vice versa. \emph{Therefore if we start with a cosine-type
integrand and perform an odd number of }$f_{z}$\emph{\ integrations, we will
end up with a sine-type output} and a change in parity of $r$. For the same
reason, since the first integrand in \eqref{f_alpha} is cosine-type with
even $r$, and case (i) can only be arrived at after an even number of $f_{z}$
\emph{\ }integrations (the number of $f_{0}$ integrations is arbitrary),
case (i)\ always involves even $r$. Repeating this reasoning explains why
case (iii) also has even $r$, while cases (ii)\ and (iv)\ have odd $r$. 

\item[5)] Dephasing results only from the terms which comprise an \emph{odd}
number of $f_{z}$ integrations. Considering that as noted in item 1) we
start from a cosine-type integral and with $r=0$, it follows from item 4)
that \emph{the last integration provides a sine-type} result.
This integral can therefore be written as a sum over terms all of which are of 
the form 
\begin{equation}
\sin [(q_{\mathrm{final}}+1+r_{\mathrm{final}}\bar{N})\theta ]\Big|_{0}^{\pi
}=0.
\end{equation}
\end{enumerate}

Recall that the operator $B_{-}$ in Eq. (\ref{Usplit}) comprises all the
terms with odd number of $\sigma _{z}\otimes B_{z}$. Hence we have proven
that the first $N$ powers in $T$ of $B_{-}(T)$ vanish, i.e., 
$B_{-}(T)=O(T^{N+1})$. This is what we set out to show and concludes the
derivation.

A remark concerning the result of Yang and Liu obtained in the interaction
picture \cite{Yang:08} is in order. They showed that $\exp(iT B_0)U(T)$
comprises only odd powers in $\sigma_z$ which are of order $T^{N+1}$ or
higher. Since $\exp(\pm iT B_0)$ does not contain any term proportional to 
$\sigma_z$ the Yang-Liu result implies our result and vice versa.

For time-dependent Hamiltonian, the proof in the interaction picture 
\cite{Pasini:09} is more convenient because powers in time occur anyway.
Therefore we stress that the statement that only odd powers in $\sigma _{z}$
of order $T^{N+1}$ or higher occur is independent of the choice of reference
frame, i.e., the description in the interaction picture or in the toggling
frame.

\section{Distance bound calculation}
\label{app:dist}

We prove the trace-norm distance bound Eq.~\eqref{D-bound-final}:

\begin{align}
& 2D[\rho _\mathrm{S}(T),\rho _\mathrm{S}^{0}(T)] \notag \\
& =\Vert \mathrm{tr}_{\mathrm{B}}\left[ \rho _{\mathrm{SB}}(T)-
\rho _{\mathrm{SB}}^{0}(T)\right] \Vert _{1} \notag \\
& =\Vert \mathrm{tr}_{\mathrm{B}}\left[ {U}(T)\rho _{\mathrm{SB}
}^{0}(0){U}^{\dag }(T)\right] 
-\mathrm{tr}_{\mathrm{B}}\left[ 
\rho_{\mathrm{SB}}^{0}(0)\right] \Vert _{1} \notag \\
& =\Vert \mathrm{tr}_{\mathrm{B}}\left[ (I\otimes B_{+}(T)+\sigma
_{z}\otimes B_{-}(T))(|\psi \rangle \langle \psi |\otimes 
\rho _{\mathrm{B}} 
\right.  \notag \\
& \quad \left. (I\otimes B_{+}^{\dag }(T)+\sigma _{z}\otimes B_{-}^{\dag
    }(T))\right] -|\psi \rangle \langle \psi |\,\Vert _{1} \notag \\
& =\Vert (b_{++}(T)-1)|\psi \rangle \langle \psi |+b_{+-}(T)|\psi \rangle
  \langle \psi |\sigma _{z}  \notag \\
& \quad +b_{-+}(T)\sigma _{z}|\psi \rangle \langle \psi |+b_{--}(T)\sigma
_{z}|\psi \rangle \langle \psi |\sigma _{z}\Vert _{1}. 
\label{D-bound1}
\end{align}
We used the definition of $b_{\alpha \beta }(T)$ 
[Eq.~(\ref{b_ab})] in
the last equality. Next, we use the triangle inequality, and 
finally the unitary invariance of the trace norm along
with the normalization of $|\psi \rangle$
\begin{align}
& 2D[\rho _\mathrm{S}(T),\rho _\mathrm{S}^{0}(T)] \notag \\
& \leq |b_{++}(T)-1|\,\Vert |\psi \rangle \langle \psi |\,\Vert
_{1}+|b_{+-}(T)|\,\Vert |\psi \rangle \langle \psi |\sigma _{z}\Vert _{1} 
\notag \\
& \quad +|b_{-+}(T)|\,\Vert \sigma _{z}|\psi \rangle \langle \psi |\,\Vert
_{1}+|b_{--}(T)|\,\Vert \sigma _{z}|\psi \rangle \langle \psi |\sigma
_{z}\Vert _{1} \notag \\
& =|b_{++}(T)-1|+|b_{+-}(T)|+|b_{-+}(T)|+|b_{--}(T)|.
\label{D-bound2}
\end{align}

\section{Correlation function inequality}
\label{app:cor-ineq}

We prove the correlation function inequality \eqref{Q}. 
Consider the spectral decomposition of the bath density operator:
$\rho_\mathrm{B}=\sum_i\lambda _{i}|i\rangle \langle i|$, where $\{
|i\rangle\}$ are normalized eigenstates, 
$\lambda_i\ge0$ are the eigenvalues, and $\sum_i\lambda_i=1$. 
Defining $|v_i\rangle := Q |i\rangle$ and $|v'_i\rangle := 
(Q')^\dag |i\rangle$, we have in this eigenbasis of $\rho_\mathrm{B}$:
\begin{align}
& \Big| \mathrm{tr}\left[ Q\rho_\mathrm{B}Q^{\prime}\right] \Big| =
\Big|\mathrm{tr}\left[ Q^{\prime}Q\rho_\mathrm{B}\right]\Big| =
\Big|\sum_i\langle i|Q^{\prime}Q|i\rangle \lambda_i\Big| \notag \\ 
&= \Big|\sum_i\langle v'_i|v_i\rangle \lambda_i\Big| \le \sum_i
|\langle v'_i|v_i\rangle| \lambda_i \le \sum_i  \| |v'_i\rangle\|
\||v_i\rangle\| \lambda_i \notag \\
& \le \sum_i\|Q^{\prime}\|\|Q\|
\lambda_i =\|Q^{\prime}\|\|Q\|
\end{align}
where we used the triangle inequality, followed by the Cauchy-Schwartz
inequality, and then the bounds
$\| |v_i\rangle\| =\|Q|i\rangle\|\le\|Q\| $
and
$\||v'_i\rangle\| =\|(Q')^\dag|i\rangle\|\le\|(Q')^\dag\| = \| Q' \|$, which
follow from the definition and properties of the
sup-operator norm (see Appendix \ref{app:norms}).


\end{document}